\documentclass[submission,copyright,creativecommons]{eptcs}
\providecommand{\event}{HCVS 2019} 
\usepackage{breakurl}             
\usepackage{underscore}           

\usepackage{xcolor}
\usepackage{xspace}
\usepackage{amsmath}
\usepackage{wrapfig}

\title{\ultimate \treeautomizer \\ 
   {\Large (\chccomp Tool Description)}}
\author{
Daniel Dietsch
\institute{University of Freiburg}
\email{dietsch@cs.uni-freiburg.de}
\and
Matthias Heizmann
\institute{University of Freiburg}
\email{heizmann@cs.uni-freiburg.de}
\and
Jochen Hoenicke
\institute{University of Freiburg}
\email{hoenicke@cs.uni-freiburg.de}
\and
Alexander Nutz
\institute{University of Freiburg}
\email{nutz@cs.uni-freiburg.de}
\and
Andreas Podelski
\institute{University of Freiburg}
\email{podelski@cs.uni-freiburg.de}
}
\def\titlerunning{\ultimate \treeautomizer}
\def\authorrunning{Dietsch, Heizmann, Hoenicke, Nutz, Podelski}

\newcommand{\initialautomaton}{\ensuremath{A_S}\xspace}
\newcommand{\ultimate}{Ultimate\xspace}
\newcommand{\treeautomizer}{TreeAutomizer\xspace}
\newcommand{\traceabstraction}{trace abstraction\xspace}
\newcommand{\chc}{CHC\xspace}
\newcommand{\chccomp}{CHC-COMP\xspace}
\newcommand{\false}{\ensuremath{\mathsf{false}}\xspace}
\newcommand{\smtlib}{SMTLIB\xspace}
\newcommand{\dofautomaton}{automaton representing the derivations of \false\xspace}

\newcommand{\ttt}[1]{\ensuremath{\texttt{#1}}\xspace}

\begin{document}
\maketitle

\begin{abstract}
 We present \ultimate \treeautomizer, a solver for satisfiability of
 sets of constrained Horn clauses.  Constrained Horn clauses (\chc) are a
 fragment of first order logic with attractive properties in terms of
 expressiveness and accessibility to algorithmic solving.
 \ultimate \treeautomizer is based on the techniques of \traceabstraction, tree
 automata and tree interpolation.
 This paper serves as a tool description for \treeautomizer in \chccomp 2019.
\end{abstract}

\section{Introduction}

We present \ultimate \treeautomizer, a solver for satisfiability of sets of
constrained Horn clauses.
The logical fragment of constrained Horn clauses (\chc) has received increasing
attention in the last years. 
One reason for its attractiveness in program verification is that it
naturally allows for expressing proof queries for many kinds of correctness
proofs, e.g., classic Floyd-Hoare proofs for while-programs, but also
assume-guarantee reasoning, compositional verification, and many
more~\cite{DBLP:conf/pldi/GrebenshchikovLPR12,DBLP:conf/popl/HoenickeMP17}.  


The \chc fragment is equivalent in expressive power to the verification of
safety properties of procedural (possibly recursive) programs, i.e., there is a
translation of a \chc-formula to a procedural program such that the \chc-formula
is satisfiable if and only if the procedural program is correct, and vice versa.
Therefore, it is not surprising that solvers for \chc-formulas often adapt
algorithms known in program verification. For example,
HSF~\cite{DBLP:conf/tacas/GrebenshchikovGLPR12,DBLP:conf/birthday/BjornerGMR15}
uses predicate abstraction, Spacer\footnote{https://spacer.bitbucket.io}
uses PDR~\cite{DBLP:conf/fmcad/EenMB11,DBLP:conf/sat/HoderB12}, and
Rahft~\cite{DBLP:conf/cav/KafleGM16} uses
\traceabstraction~\cite{DBLP:conf/sas/HeizmannHP09}, 
to name just a few tools. 
\ultimate \treeautomizer is part of this tradition and is an adaptation of the
\traceabstraction verification algorithm for procedural 
programs~\cite{DBLP:conf/popl/HeizmannHP10}.  

This paper is a tool description for the \treeautomizer tool as it participated
in \chccomp in 2018 and 2019.  
We give a brief overview of how \traceabstraction is used to solve \chc-formulas.
Afterwards, we describe some aspects of the implementation of \treeautomizer and
some crucial optimizations.
Last, we discuss expected strengths and weaknesses of the approach.

\goodbreak

\section{Approach}

In this section, we describe the approach for solving formulas in the
\chc-fragment used in \treeautomizer. 
The approach is based on \traceabstraction~\cite{DBLP:conf/sas/HeizmannHP09};
its adaptation to solving \chc-formulas has been described by Kafle and
Gallagher~\cite{DBLP:conf/vmcai/KafleG15} and by Wang and
Jiao~\cite{DBLP:journals/cj/WangJ16}, we refer to these papers for a more
in-depth description and only give an overview here.

In the following, we assume that a constraint theory $T$ is given, and that we
have an SMT-solver for $T$. 
Furthermore, we refer to constraints over theory $T$ with free variables
$\vec{x}$ as $C(\vec{x})$ and we assume that a set $\{ P_1, P_2, \ldots \}$ of
predicate symbols is given that are not used by the constraint theory $T$.

A formula in the \chc-fragment is given as a set of clauses where each clause
is of one of the below forms. According to general convention, Horn clauses
subdivided the categories of \emph{facts}, 
\emph{definite clauses}, and
\emph{queries} (also: \emph{goal clauses}), depending on which of the below patterns they match.

\begin{align*}
  \forall \vec{x} \ldotp  C(\vec{x}) \to & P(\vec{x}) & & \text{(fact)} \\
  \forall \vec{x} \ldotp P_1(\vec{x}) \land \ldots \land P_n(\vec{x})
  \land C(\vec{x}) \to & P(\vec{x}) & & \text{(definite clause)} \\ 
  \forall \vec{x} \ldotp P_1(\vec{x}) \land \ldots \land P_n(\vec{x})
  \land C(\vec{x}) \to & \false & & \text{(query)} 
\end{align*}
In the remainder, we assume that a set $S$ of constrained Horn clauses is given.

Now, let us consider the resolution trees over the clauses in set $S$ with root \false.  We
call such a tree a 
\emph{derivation of \false}.
Since no constraints ever occur in a clause head, the resolvent at the root
of a derivation of \false is a query with one large conjunctive constraint in the antecedent,
i.e., it is of the following form.
$$\forall \vec{x} \ldotp C_1(\vec{x}) \land \ldots \land C_n(\vec{x}) \to \false$$

We call a derivation of \false \emph{feasible} if the formula $\exists \vec{x} \ldotp C_1(\vec{x}) \land \ldots \land C_n(\vec{x})$ is satisfiable, and \emph{infeasible} otherwise.
The existence of a feasible derivation of \false means that the
conjunction of the clauses in $S$ is contradictory. Completeness of first-order resolution implies that the converse
also holds, i.e., that the absence of a feasible derivation of \false implies
satisfiability of the formula.
Thus, we can formulate the following proof rule.

\begin{quote}
 A set of constraint Horn clauses $S$ is satisfiable if and only if there is no
 feasible derivation of $\false$ over $S$.
\end{quote}

\ultimate \treeautomizer's approach to prove satisfiability of the set of Horn
clauses $S$ is to show infeasibility of all derivations of \false over $S$. 
The refinement algorithm used for this purpose is shown in
Figure~\ref{fig-refinement}.
The proving process starts by sampling a derivation from the set of
all derivations of \false over $S$. The sample derivation is then checked for
feasibility using an SMT solver. If the sample derivation is feasible, the clause
set $S$ is unsatisfiable (since it implies $\false$).
If the sample derivation is infeasible, the sample is generalized to a set of
derivations which are all infeasible. This set is subtracted from the set of
derivations of \false. 
This process is repeated until either all derivations of \false have been proven
infeasible or a feasible derivation has been found.

\begin{wrapfigure}{R}{0.5\textwidth}
  \centering
  \begin{minipage}{5cm}
\begin{verbatim}
A := A_S
while (nonempty(A)) {
  d := sample(A)
  if (d is feasible)
    return unsat
  I := getTreeInterpolant(d)
  G := generalize(d, I, S)
  A := A \ G
}
return sat
\end{verbatim}
  \end{minipage}
  \caption{Trace abstraction refinement scheme used in
  \ultimate \treeautomizer. \ttt{S} is the input set of constrained Horn
  clauses. \ttt{A_S} is a set containing all derivations of \false over $S$. The
  procedure \ttt{sample} picks an element from a non-empty set. The procedure
  \ttt{generalize} takes an infeasible derivation of \false as input and returns
  a set of infeasible derivations of \false that contains at least the input
  derivation (see also Figure~\ref{fig-generalize}). Note that the check for
  feasibility as well as the \ttt{generalize} procedure rely on calls to an
  (interpolating) SMT solver.\label{fig-refinement}}
\end{wrapfigure}





\newpage

\section{Implementation}

\treeautomizer is implemented in the \ultimate framework. 
It is written in Java, open source, and can be downloaded and contributed to on
\ultimate's Github page\footnote{\url{https://github.com/ultimate-pa/}}.

\ultimate provides for \treeautomizer the \smtlib parser, utilities for
handling formulas (e.g., simplifications), and the \ultimate Automata Library.
SMT solvers for which by \ultimate provides an interface include
SMTInterpol~\cite{DBLP:conf/spin/ChristHN12}, 
Z3~\cite{DBLP:conf/tacas/MouraB08}, 
CVC4~\cite{DBLP:conf/cav/BarrettCDHJKRT11}, and
MathSat~\cite{DBLP:conf/tacas/CimattiGSS13}.  In cases when a solver does not support
interpolation in the given constraint theory, but can produce unsatisfiable
cores, Newton-style interpolation~\cite{DBLP:conf/sigsoft/DietschHMNP17} can be used
to obtain interpolants.
 


\treeautomizer takes as input Horn clause sets in the format used in 
\chccomp\footnote{\url{https://chc-comp.github.io/2018/format.html}}.
During parsing, the input formulas are converted into the normal form given
above.

In order to represent (possibly infinite) sets of derivations of \false,
\treeautomizer uses tree automata (see \cite{tata2007} for more details on tree automata). 
The alphabet that the tree automata operate on is the set of input Horn clauses $S$.
The states of the tree automata have one of two different semantics.  
The states of the automata \ttt{A}, and \initialautomaton represent the
uninterpreted predicates in the set  $\{P_1, P_2 \ldots \}$.
The states of the interpolant automaton \ttt{G} represent the interpolants from
the interpolation query that is generated from the sample derivation of \false
\ttt{d}.
From this sample query, a generalization procedure computes the canonical
interpolant automaton.  
The canonical interpolant automaton is given by the set of all rules that
correspond to a valid implication between the formulas in the source of the
automaton rules, the constraint in the alphabet symbol, and the formula at target
of the rule (see \cite{DBLP:journals/cj/WangJ16}) for a thorough description).

\section{Optimizations}

In each iteration of the main refinement loop of \traceabstraction, an
interpolant automaton (\ttt{G}) is created and subtracted from the
\dofautomaton (\ttt{A}). Two major bottlenecks in terms of space and time
consumption may arise from this.
First, the generalization that is done during creation of the interpolant
automaton can produce a large number of candidate transition rules each one
requiring an SMT solver call.
Second, the difference operation requires construction of a product automaton
and thus can lead to growth of the \dofautomaton that is exponential in the
number of loop iterations.
Both problems are amplified by an increasing nonlinearity of the involved Horn
clauses. 

\paragraph{Minimization}

The explosion through the repeated product construction can often be contained
through an additional minimization step on the result of the difference
operation. Standard minimization algorithms for tree automata can be used here;
the \ultimate automata library currently supports two minimization variants, one
based on the naive algorithm~\cite{tata2007} the other on bisimulation~\cite{DBLP:conf/wia/AbdullaKH06}.

\begin{wrapfigure}{R}{0.55\textwidth}
\begin{verbatim}
generalize(d, I, S) {
  result = freshTreeAutomaton(S);
  for ((P_1 ... P_n /\ C -> P) in S) {
    for (formulas (phi_1, ..., phi_(n+1)) 
       with phi_i occur in I) {
      candidateRule := 
        phi_1 /\ ... phi_n /\ c -> phi_n+1;
      if (checkvalidity(candidateRule))
        result.addRule(candidateRule);
    }
  }
  return result;
}
\end{verbatim}
  \caption{\ttt{generalize} procedure as called in the refinement algorithm in
  Figure~\ref{fig-refinement}. 
  Given an infeasible derivation of \false \ttt{d}, a tree interpolant $I$,
  and the Horn clause set \ttt{S}, the procedure returns a tree automaton that
  accepts \ttt{d} and, by generalization, possibly other infeasible derivations
  of \false.  The tree automaton's states are the predicates that occur in the
  tree interpolant \ttt{I}.
  The generalization happens through adding rules to the tree automaton
  that correspond to a valid implication between the source states, conjoined
  with the constraint in the alphabet symbol (a Horn clause from $S$), and the
  target state.
  \label{fig-generalize}}
\end{wrapfigure}

\paragraph{On-Demand Construction of Interpolant Automaton}

The explosion of the number of rules in the interpolant automaton can be
countered by integrating the difference (i.e., complementation and
intersection) operation with the creation of the interpolant automaton.

The idea behind the integration is that a large number of candidate rules in the
interpolant automaton is irrelevant to the result of the difference operation. 
The basic intuition here is that for computing the difference $S\setminus T$ of 
two sets $S$ and $T$, only the part of $T$ that lies in the intersection of $S$
and $T$ is relevant -- elements of $T$ that don't lie in $S$ need not be 
considered by a subtraction algorithm.
For the subtraction of tree automata, this means the following:
A rule is irrelevant to the result of the difference operation, 
$\ttt{A} - \ttt{G}$, if it never contributes to the construction of a tree in
\ttt{G} that lies in the language of \ttt{A}. Such candidate rules can be
filtered out during the product construction by only querying the interpolant
automaton for rules whose source tuple is reachable in according to the minuend
automaton (\ttt{A}).

\goodbreak

\section{Discussion}

\treeautomizer's approach inherits its basic properties from the
\traceabstraction approach. 
Thus, \treeautomizer is conceptually sound and relatively complete. As is
common for such refinement schemes, the actual detection of a proof of
satisfiability (and thus actual completeness) depends on guessing the right
formulas during the generalization step (we only mentioned interpolation here,
but several other methods are available).

We believe that one strength of the \traceabstraction approach lies in
a semantic independence of refinement steps. 
For example in predicate abstraction with
CEGAR~\cite{DBLP:conf/cav/ClarkeGJLV00} (which several program verification
schemes can be seen as a variant of), formulas that stem from many different
refinement steps are conjoined. This means that SMT-solver
queries get bigger and bigger over a growing number of iterations, which can
swamp the SMT solver. 
In \traceabstraction on the other hand, the formulas used in the
\ttt{generalize} procedure can be forgotten, after the difference \ttt{A - G}
has been computed, i.e. formulas from different refinement steps are never
conjoined.
However, among other things, this property relies on a rich-enough structure of
the initial automaton. In particular, this means that \chc-formulas that stem
from proof queries for programs where large block encoding has been performed or
where the program counter is not made explicit by using different uninterpreted
predicates for each location, this compositionality may not come into full
effect.


 \bibliographystyle{eptcs}
 \bibliography{main}

\end{document}